\documentstyle[twocolumn,aps]{revtex}
\begin{document}
\draft

\title{The structure of large $^3$He-$^4$He mixed drops
around a dopant molecule}

\author{Mart\'{\i} Pi, Ricardo Mayol and Manuel Barranco}

\address{
Departament ECM,  Facultat de F\'{\i}sica,
Universitat de Barcelona.
E-08028 Barcelona, Spain}

\date{\today}

\maketitle

\begin{abstract}

We have investigated how helium atoms are distributed within a mixed
$^3$He$_{N_3}$-$^4$He$_{N_4}$ large drop with $N_3 \gg N_4$.
For drops  doped with a SF$_6$ molecule or a Xe atom, we have found 
that the number of $^3$He atoms within the volume containing 
the first two solvation shells increases when $N_4$ decreases 
in  a way  such that these dopants may be in a superfluid
environment for $N_4 \geq$ 60, which gradually disappears as $N_4$
decreases. The result is in qualitative agreement with recent
experimental data.

\end{abstract}


In a recent experiment, Grebenev et al \cite{Gr98} have carried out
the equivalent of the Andronikashvili experiment \cite{An46}
in a microscopic system, namelly a mixed $^3$He-$^4$He drop consisting
of about 10$^4$ atoms doped with an oxigen carbon sulfide (OCS)
molecule. By analyzing the infrared spectrum of OCS, Grebenev et al
(see also Ref. \onlinecite{Le98}) conclude that the molecule freely rotates
when a number of $^4$He atoms large enough  coats  the impurity,
preventing
the $^3$He atoms, which are in the normal phase at a temperature of the
order of 150 mK \cite{Br90,Gu91}, from getting too close to the OCS
molecule. That number is of the orden of 60, in excelent agreement
with path integral \cite{Si89} and variational \cite{Ra90}  
Monte Carlo calculations. It is
remarkable that the presence of the impurity, which causes the $^4$He
density to rise up to several times the saturation value,
is not destroying its superfluid character, and that in spite of the 
high densities reached, the first solvation shell remains 
liquid\cite{Kw96}. An indication of this fluidlike behavior is that 
the peak density in the first solvation shell continues to increase
as the second shell grows\cite{Bl96}.

Even if the intrepretation of the microscopic Andronikashvili
experiment is on a firm basis, a remaining major question is how
$^3$He is distributed around the $^4$He-plus-impurity complex, and
in general, how liquid $^3$He is dissolved into $^4$He droplets
at very low temperatures. These are the questions we
want to address in this work.

At zero temperature, it is known
that the maximum solubility of $^3$He in the bulk of $^4$He is $\sim
6.6 \%$ \cite{Ed92}. For liquid $^4$He systems having a
free surface, it is also known that a large amount of $^3$He
is accumulated on the free surface occupying  Andreev
states \cite{An66,Ed78} before it starts being dissolved into the
bulk. In the case of drops
made of up to several thousands of atoms, the surface region constitutes
a sizeable part of the system \cite{Ha98}, and the surface has
a large capacity of storing $^3$He atoms before they get inside the
drop \cite{Ba97}. Due to the wide free surface
of both isotopes \cite{St87a,Lu93} and to the low surface
tension
of the $^3$He-$^4$He liquid interface \cite{Sa97}, one  expects that
this region plays a prominent role when it constitutes a large part
of the system or, as in the present case, when it is
close to the foreing atom or molecule.

The structure and energetics of mixed, doped or not, helium droplets
has been  addressed using a finite-range density functional\cite{Ba97}.
 That work was carried out before the experiments
reported in Ref. \onlinecite{Gr98}, and the emphasis was put on improving
the density functional  to better describe the
thermodynamical properties of the liquid mixture, and to study
rather small mixed droplets with $N_4 \gg N_3$. Our main  goal here is
to apply the density functional method to droplets whose
characteristics are closer to those of the experiments, with the
restriction of spherical symmetry for the He-impurity potential
for the sake of simplicity. We
have considered Xe and SF$_6$ as dopants, using for the later a
spherically averaged interaction potential. The Xe-He potential is
weaker than the SF$_6$-He one. In this respect, our results for that
atomic impurity  should better represent the experimental ones for OCS
even if this linear molecule produces deformations in the helium drop
that we have not considered here.
The density functional and the treatment of the impurity are 
thoroughly described in Ref. \onlinecite{Ba97}.

The large number of $^3$He atoms in the droplets we are describing
($N_3 > 1000$) allows us to employ an Extended
Thomas-Fermi method to describe the fermionic component
of the mixture. We have used for the $^3$He kinetic energy density
the expression given in Ref. \onlinecite{St87}, which contains up to second
order density gradient corrections to the standard
$\sim \rho_3^{5/3}(r)$ expression, where $\rho_3$
and $\rho_4$ will denote the particle density of each isotope.
We have checked that this density functional reproduces
accurately the Hartree-Fock results\cite{Ba97} obtained for the
largest drops there studied (see also Refs. \onlinecite{Gu91,St87}).

Figure \ref{fig1} displays the situation in which a $^4$He$_{728}$
drop, whose size is large enough to clearly distinguish in it a
surface and a bulk region,
is coated with an increasing number of $^3$He atoms, and the limiting
situation of the same drop immersed into liquid $^3$He. The evolution
with $N_3$ of the $^3$He concentration inside the  $^4$He drop,
defined as $x_3 \equiv \rho_3/(\rho_4 + \rho_3)|_{bulk}$, is
shown in the insert. Several interesting features of this figure
are worth to comment. First of all, a fairly large amount of $^3$He is
needed before it is appreciably dissolved in the bulk:
for $N_3 = 1000$, $\rho_3$ near the origin is $\sim 1.4 \times
10^{-8} {\rm \AA}^{-3}$. The solubility is appreciably reduced
by finite size effects. Indeed, one can see from the insert that
the limiting solubility into the $N_4$ = 728 drop
is $\sim 2.5 \%$, as compared to the
6.6 \% value in the liquid mixture. It is also worth to see that
for large $N_3$ droplets, the bulk solubility is slightly higher
than the limiting solubility, indicating
that finite size effects still appear in rather
large drops. Another manifestation of a finite size effect is that the
average $^3$He density is above the saturation value even for the
larger drops, showing that the existence
of the outer $^3$He surface still causes a visible
density compression.

Due to the high incompressibility
of helium, the bulk density of $^4$He {\it decreases} when $^3$He is
dissolved, and the rms radius of the $^4$He drop manifests a peculiar
$N_3$ behavior. It decreases when $N_3$ increases up to a few
hundreds due to the initial
compression of the outermost $^4$He surface, and then
steady increases as $^4$He is pushed off the center by intruder
$^3$He atoms. This is a very tiny effect anyway. For example, we have
found that the rms radius of the  $^4$He$_{728}$ drop
is 15.70 {\rm \AA}. It decreases when $^3$He is added,
reaching a minimum value of  15.64 {\rm\AA} for $N_3 \sim$ 250, and
then it steadily increases up to 16.11 {\rm \AA} for $N_3$ = 10000.
The rms radius of the  $^4$He$_{728}$ drop immersed into liquid
$^3$He is 16.14 {\rm \AA}.

When a SF$_6$ molecule is captured by a helium drop, it moves into the
bulk producing a drastic rearrangement of the drop density around
it\cite{Ba93,Ch95,Ga97,Ga98}.  For large  
$^4$He droplets, it is especially noteworthy the appearence of two
high density solvation shells with a density-depleted
region in between. It is then natural to  ask about the
possible existence of Andreev-like states arising at the `inner
$^4$He surface', and  whether a large number of
$^3$He atoms can be stored there, producing an onion-like structure
of alternative $^4$He and $^3$He shells around the impurity, or even
if when $N_3 \gg N_4$, the later can displace the former in the first
solvation shell.

Figure \ref{fig2} shows the density profiles of several
$^4$He$_{728}\, + \,^3$He$_{N_3} + $ SF$_6$ droplets, giving a
positive
answer to the first question, and a negative answer to the other two.
We have found that indeed, about {\it one} $^3$He atom is in the inner
surface, but that $^3$He is mostly coating the $^4$He-plus-impurity
complex, as in undoped droplets. To check this result we have
started the calculations from different initial shapes,
some having the
onion-like form mentioned before. It has turned out that these
always are high energy, metastable configurations,
and the mixed droplet eventually evolves towards stable configurations
of the kind shown in Fig. \ref{fig2}. The larger zero point motion
energy of $^3$He makes it energetically more advantageous to fill the 
first solvation shell with $^4$He atoms, and $^3$He is expelled to
the outer region of the drop.

We are now in position to discuss  a physical situation relevant to
the microscopic Andronikashvili experiment. We first observe that
the first solvation shell \cite{note} can host
$\sim$ 23 $^4$He atoms in the case of SF$_6$ as a dopant, and $\sim$
15 atoms in the case of Xe\cite{Ba93,Ch95,Ga97}.
According to Refs. \onlinecite{Si89,Ra90,Kw96},
these numbers are too small for the $^4$He droplet being superfluid.
It is thus crucial to know how the second solvation shell is built,
especially what is its composition. Too many $^3$He atoms in that shell
might shrink or even wash out the superfluid environment around the
dopant.
The density functional method cannot tell whether a given
configuration is superfluid or not, but it can give a quantitative
answer to its local composition because it is able to reproduce 
available microscopic density profiles\cite{Ba93,Ch95,Pa86}. We present 
examples of such compositions in Figs. \ref{fig3} and \ref{fig4}.

Figure \ref{fig3} shows the density profiles for
$^4$He$_{N_4} \,+\, ^3$He$_{1000}$ + SF$_6$ and
$^4$He$_{N_4}\, +\, ^3$He$_{1000}$ + Xe
with $N_4$ = 35, 60 and 100.
We have carried out calculations for two different dopants to see
what is the influence of the He-impurity potential on the results. It
turns out that a weaker attractive potential favors the mixing of both
isotopes in the whole allowed volume (the Xe-He and SF$_6$-He
potentials are plotted in Ref. \onlinecite{Ga97}, for instance).
However, this is in part a first glance
effect, since the number of $^3$He atoms in the first solvation shell
around Xe is less than one (see Fig. \ref{fig4}). Rather, the relevance
of Fig. \ref{fig3} lies in that it  shows
how $^3$He is filling the second solvation shell as $N_4$ decreases.

A more quantitative look at this phenomenon is presented  in Fig.
\ref{fig4}, where
we have plotted the number of atoms of each isotope as a function of
the radial distance to the center of the drop. Notice 
 that for a given impurity, the number of $^4$He atoms in the first
solvation shell (extending up to $\sim 5.5 \,{\rm \AA}$) is sensibly
the same for the three selected $N_4$ values.
It is also  worth to look at the ratios $N_3/(N_4 + N_3)$
within the second solvation shell which extends from $\sim 5.5$ to
$\sim 8.5\, {\rm \AA}$. In the SF$_6$ case, they
are $\sim 8 \%$ for $N_4$ = 100,
$\sim 29 \%$ for $N_4$ = 60, and $\sim 65 \%$ for $N_4$ = 35.
Considering  the content of the  two shells, these ratios are
$\sim 5 \%$, $\sim 19 \%$, and $\sim 41 \%$ which correspond,
respectively, to 3, 10, and 22 $^3$He atoms.
The values for Xe are slightly smaller.
These numbers make it quite plausible that a SF$_6$  molecule
or a Xe atom
in a $^4$He$_{N_4} \,+\, ^3$He$_{1000}$ drop is in a superfluid
environment when $N_4$ = 100 or 60, 
whereas it is not when $N_4$ = 35, as the microscopic
Andronikashvili experiment indicates for OCS. 

We are indebted to Peter Toennies and Andrej Vilesov for useful
discussions. This work has been performed under grants
PB95-1249 and PB95-0271-C02-01 from CICYT, Spain
and Program 1996SGR-00043 from Generalitat of Catalunya.

\begin{figure}
\caption{ Density profiles of  $^4$He$_{728}\, +\, ^3$He$_{N_3}$
droplets for $N_3$ values from 1000 to 10000 in $\Delta N_3$ = 1000
steps. To ease the figure, only the $^4$He densities corresponding
to  a few $N_3$ cases have been plotted. Also shown is the density
profile of a $^4$He$_{728}$ drop immersed into liquid $^3$He
(dotted lines). Insert: bulk $^3$He concentrations.
The connecting solid line is to guide the eye.
Also shown is the value corresponding to $^4$He$_{728}$ in liquid
$^3$He (dotted line).
}
\label{fig1}
\end{figure}
\begin{figure}
\caption{ Density profiles of  $^4$He$_{728}\, +\, ^3$He$_{N_3} +
$ SF$_6$ droplets for for $N_3$ values from 4000 to 10000
in $\Delta N_3$ = 1000 steps.
 }
\label{fig2}
\end{figure}
\begin{figure}
\caption{ Bottom panel: Density profiles of
$^4$He$_{N_4}\, + \,$  $^3$He$_{1000}\, + $
  SF$_6$  droplets for $N_4$ = 35, 60 and 100.
Top panel: Density profiles of
$^4$He$_{N_4}\, +\,$ $^3$He$_{1000}$ + Xe
droplets for the same $N_4$ values.
 }
\label{fig3}
\end{figure}
\begin{figure}
\caption{ Top panel: Number of $^4$He atoms as a function of the
radial distance for the droplets of Fig. 3.
Bottom panel: Number of $^3$He atoms
as a function of the radial distance for the same droplets.
 }
\label{fig4}
\end{figure}
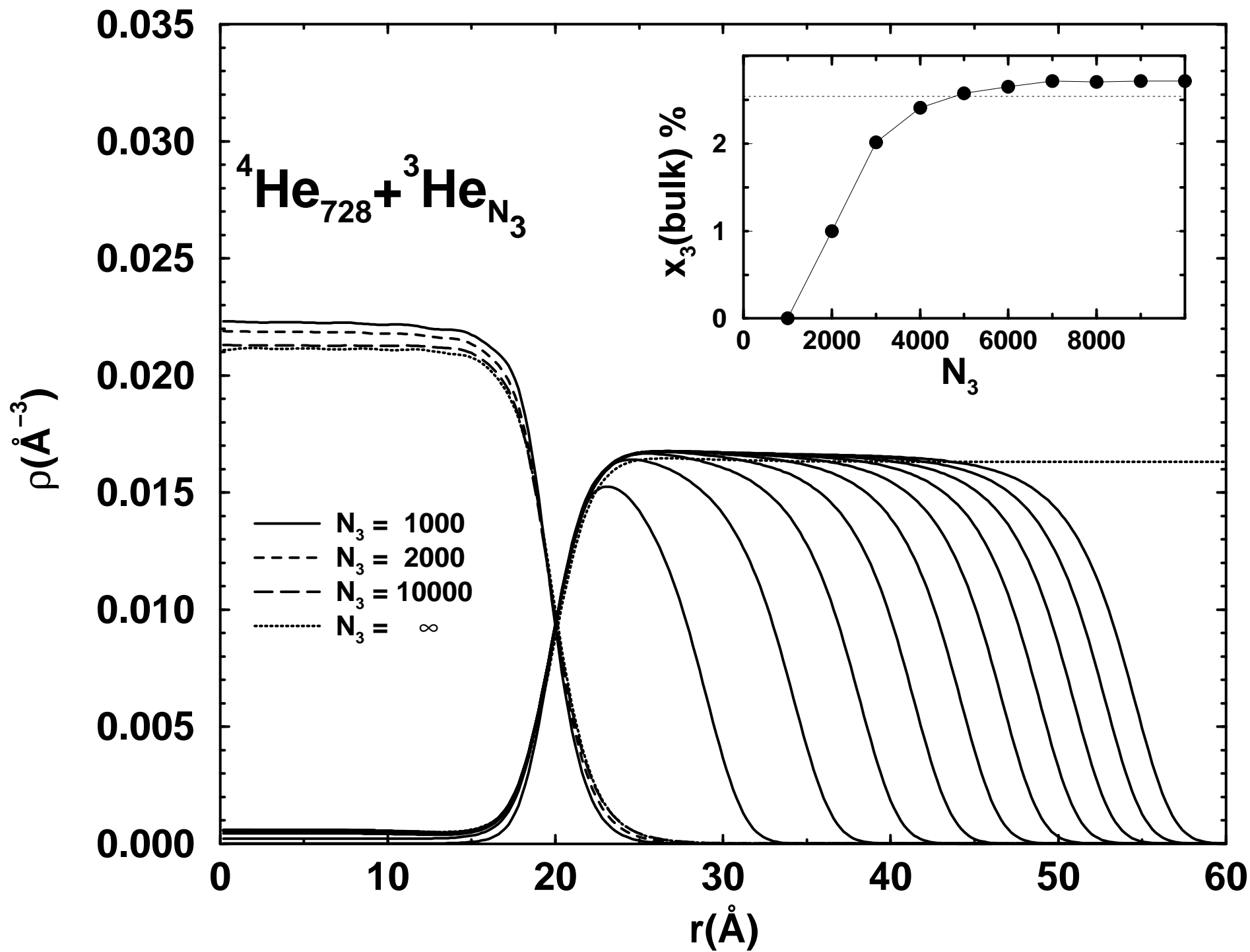
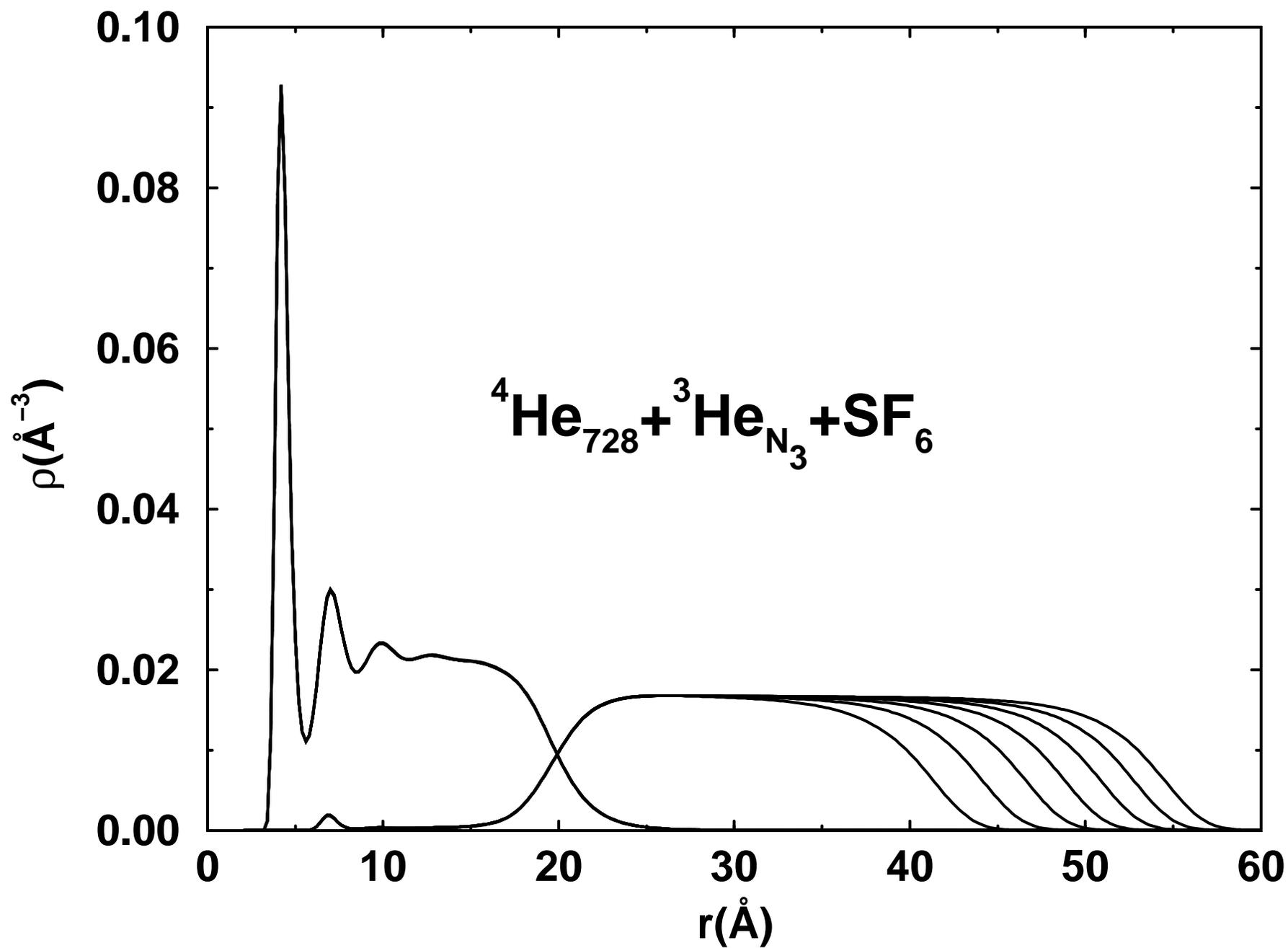
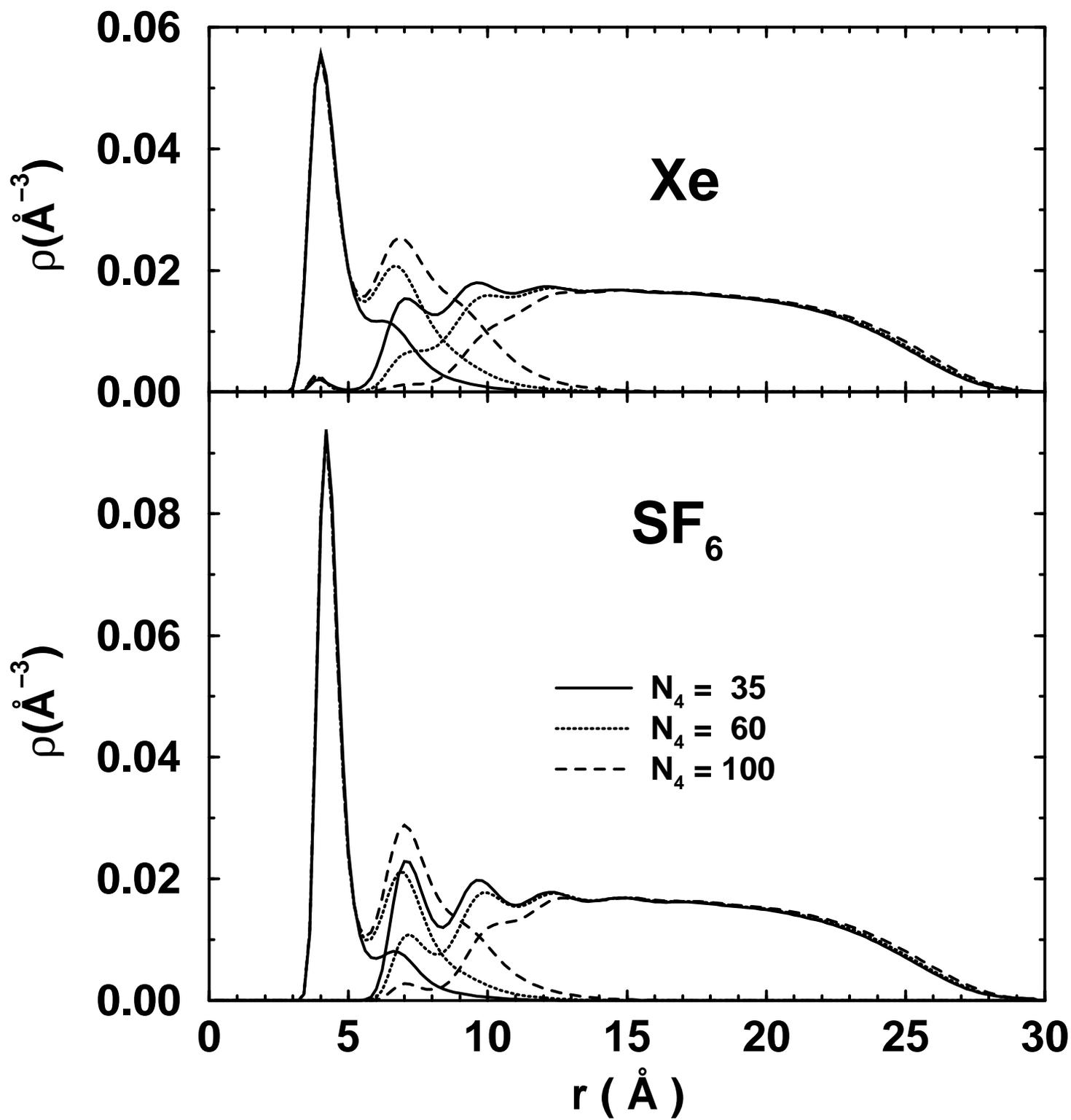
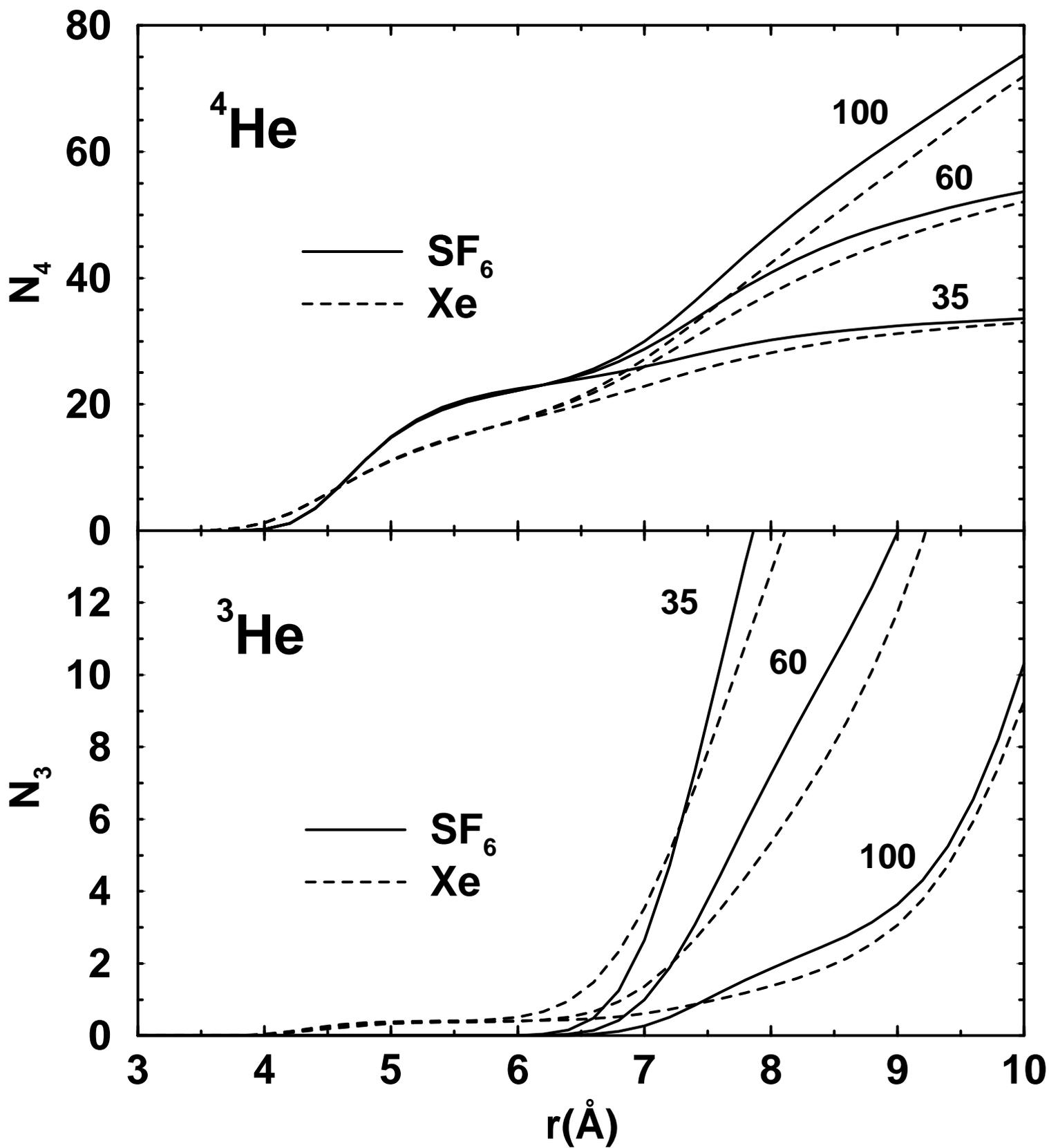
\end{document}